\documentclass[twocolumn ,amssymb,amsmath,nobibnotes,aps]{revtex4}
\usepackage{graphicx}
\usepackage{dcolumn}
\usepackage{bm}

\def\be{\begin{equation}}
\def\ee{\end{equation}}
\def\bea{\begin{eqnarray}}
\def\eea{\end{eqnarray}}

\begin{document}

\title{Hadrons as Skyrmions in the presence of isospin
chemical potential}
\author{M. Loewe}
\email{mloewe@fis.puc.cl} \affiliation{Facultad de F\'\i sica,
Pontificia Universidad Cat\'olica de Chile,\\ Casilla 306, Santiago
22, Chile.}
\author{S. Mendizabal}
\email{smendiza@fis.puc.cl} \affiliation{Facultad de F\'\i sica,
Pontificia Universidad Cat\'olica de Chile,\\ Casilla 306,
Santiago 22, Chile.}
\author{J.C. Rojas}
\email{jurojas@ucn.cl} \affiliation{Departamento de F\'{\i}sica,
Universidad Cat\'{o}lica del Norte,\\ Casilla 1280, Antofagasta,
Chile}

\begin{abstract}
The stability of the Skyrmion solution in the presence of finite
isospin chemical potential $\mu $ is considered, showing the
existence of a critical value $\mu _{c} = 222.8$ MeV where the
Skyrmion mass vanishes. Using the Hamiltonian formulation, in terms
of collective variables, we discuss the behavior of different
skyrmionic parameters as function of the isospin chemical potential
($\mu$), such as the energy density, the isoscalar radius and the
isoscalar magnetic radius. We found that the radii start to grow
very fast for $\mu \geq 140$ MeV, suggesting the occurrence of a
phase transition.
\end{abstract}

\maketitle

The skyrmion picture has attracted the attention of many authors as
 a possible way for understanding hadronic dynamics and the hadronic
 phase structure. The behavior of hadrons in a media,
i.e. taking into account temperature and/or density effects, can be
analyzed according to this perspective.

 The Skyrme lagrangian is

\bea \label{lagrangian}{\cal L} &=& \frac{F_{\pi}^2}{16} Tr\left[
\partial_{\mu}U \partial^{\mu}U^{\dagger}
\right] \nonumber \\ &&+ \frac{1}{32e^2}Tr \left[
(\partial_{\mu}U)U^{\dagger}, (\partial_{\nu}U)U^{\dagger}
\right]^2, \eea

\noindent where $F_{\pi}$ is the pion decay constant and $e$ is a
numerical parameter. The isospin chemical potential is introduced as
a covariant derivative of the form \cite{actor}

 \be
\partial_{\nu} U \rightarrow  D_{\nu}U = \partial_{\nu}U -i \frac{\mu}{2} [\sigma^3,U] g_{\nu
0}. \ee

 The $U$ field matrix, for the static case, can be parameterized in the standard way

\be U=U_0 = \exp \left( -i\xi \vec{\sigma}\cdot \hat{n}
\right)=\cos\xi -i(\vec{\sigma} \cdot \hat{n})\sin\xi, \ee

\noindent where $\vec{\sigma}$ is the sigma matrix vector and
$\hat{n}^2=1$. This ansatz has a ``Hedgehog" shape obeying the
boundary conditions \cite{ad-witt}

\bea &&\xi(\vec{r})=\xi(r), \; \; \hat{n}=\hat{r},
\nonumber \\
&&\xi(0)=\pi, \; \; \xi(\infty)=0. \eea

The mass of the Skyrmion, for static solutions, develops a
dependence on the Isospin Chemical potential as well as on the
temperature.  Defining  $\hat{r}=e F_{\pi} r$, the mass of the
Skyrmion will be given by

\be M_{\mu}=M_{\mu=0}- \frac{\mu^2}{4 e^3 F_{\pi}} I_2 -
\frac{\mu^2}{32 e^3 F_{\pi}} I_4, \label{mmu}\ee

\noindent where $M_{\mu=0}$ is the zero chemical potential
contribution. Notice that the chemical potential terms contribute
with opposite sign. This implies that the solution will become
unstable above certain value of $\mu$. In the previous equation

\bea M_{\mu=0} &=& \frac{F_{\pi}}{4 e} \left\{ 4 \pi
\int_0^{\infty}d\hat{r} \left[ \frac{\hat{r}^2}{2} \left(
\frac{d\xi_1}{d\hat{r}}\right)^2 + \sin^2(\xi_1)\right]\right.
\nonumber
\\  && + 4 \pi \int_0^{\infty}d\hat{r}
\frac{\sin^2(\xi_1)}{\hat{r}^2} \times \nonumber \\ & &  \left.
\left[ 4\hat{r}^2 \left(
\frac{d\xi_1}{d\hat{r}}\right)^2+2\sin^2(\xi_1)\right] \right\}.\eea

Assuming a radial profile ($\xi=\xi(r)$, the integrals $I_2$ and
$I_4$ are given by

\bea I_2 &=& \frac{4 \pi}{3} \int d\hat{r} \hat{r}^2 \sin^2 \xi,\\
I_4 &=& \frac{32 \pi}{3} \int d\hat{r} \hat{r}^2 \left[ \sin^2 \xi
\left( \frac{d\xi}{d\hat{r}}\right)^2+ \frac{4}{\hat{r}^2} \sin^4
\xi \right].\nonumber \label{integrals}\eea

\begin{figure}
\includegraphics[angle=0,width=0.42\textwidth]{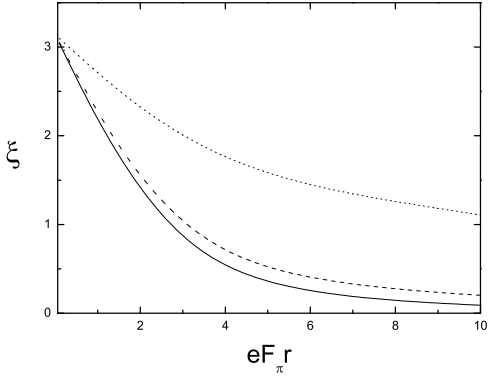}
\caption{\label{perfilvsradio} Numerical solution for the profile
$\xi(r)$ for different values of $\mu$. ($\mu=12.9$ (MeV): solid
line; $\mu=25.8$ (MeV): dashed line; $\mu=38.7$ (MeV): dotted
line.)}
\end{figure}

In order to minimize the mass, we use a variational procedure which
leads us to the following condition for the radial profile

\begin{eqnarray}
&&\left(\frac{1}{4}\hat{r}^2+2\sin^{2}{\xi}\right)\frac{d^2\xi}{d\hat{r}^2}+\frac{1}{2}\hat{r}\frac{d\xi}{d\hat{r}}+
\sin{2\xi}\left(\frac{d\xi}{d\hat{r}}\right)^{2}\nonumber\\
&&-\frac{1}{4}\sin{2\xi}-\frac{\sin^{2}
{\xi}\sin{2\xi}}{\hat{r}^{2}}\nonumber\\&&
-\frac{\hat{\mu}^{2}\hat{r}^{2}\sin^{2}{\xi}}{3}\left(\frac{1}{2}\frac{d^2\xi}{d\hat{r}^2}
+\frac{1}{2\sin{\xi}}\left(\frac{d\xi}{d\hat{r}}\right)^{2}\right.\nonumber\\
&&+\left.
\frac{1}{\hat{r}}\frac{d\xi}{d\hat{r}}-\frac{1}{4}\frac{\sin{2\xi}}{\sin^2{\xi}}-\frac{2\sin{2\xi}}{\hat{r}}\right)=0,
\label{ecmovi}
\end{eqnarray}

\noindent where $\hat{\mu}=\mu/(eF_{\pi})$. Equation (\ref{ecmovi})
can be solved numerically for the profile $\xi(r)$, for different
values of $\mu$ (Figure \ref{perfilvsradio}), notice that the width
of the profile grows with the chemical potential.

\begin{figure}
\includegraphics[angle=0,width=0.45\textwidth]{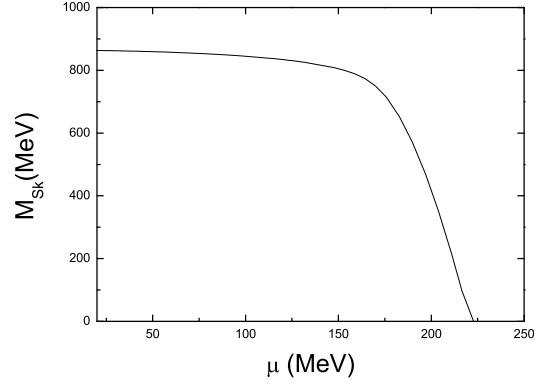}
\caption{\label{numericomasa} The numerical solution for the
Skyrmion mass as funtion of $\mu$.}
\end{figure}

In order to obtain the mass of the Skyrmion, the profile has to be
inserted, numerically, in equation (\ref{mmu}). Figure
\ref{numericomasa} shows the chemical potential dependence of the
mass. The point where the mass vanishes corresponds to the critical
value $\mu_c=222.8$ MeV. This value does not depend on a model for
the shape of $\xi(r)$. It is a fundamental result associated to the
Skyrmion picture.

In \cite{lmr2}, a discussion was presented about the validity of the
Atiyah-Manton ansatz \cite{at-man} at finite chemical potential. It
turns out that this procedure is limited only for small values of
($\mu<$100 MeV), been in this region quite in agreement with the
numerical solution.

The Skyrmion solution we have presented, can be considered as the
basic or skeleton structure for the hadronic states. In order to
characterize different hadronic states, and following
\cite{ad-witt}, we will introduce collective variables $A(t)$,
computing then the mass spectrum of the nucleons, in an hadronic
approach. As usual, the $SU(2)$ collective coordinates $A(t)$ are
introduced as

\be U=A(t) U_0 A^{\dag}(t).\label{A}\ee

\noindent where

\be A(t)=a_0(t) \sigma_0+i\vec{a}(t) \cdot \vec{\sigma},
\label{A2}\ee

\noindent where the $a$'s obey the constraint

\be a_0^2(t)+\vec{a}^2(t)=1.\ee

Introducing (\ref{A}) and (\ref{A2}) into (\ref{lagrangian}), a
rather length computation leads us to the Lagrangian

\bea L = -M_{\mu} + 2\lambda
\left(\dot{a}_i+\mu\frac{\tilde{A}_i}{2} \right)^2,
\label{lagrangian}\eea

\noindent where $\lambda=\left(2 \pi/3e^3 F_{\pi} \right) \Lambda$,
with

\be \Lambda= \int r^2 \sin^2F\left[1+4 \left(
F'^2+\frac{\sin^2F}{r^2} \right) \right]. \label{Lambda}\ee

\noindent In equation (\ref{lagrangian}) we have defined
$\tilde{A}_i =  g_{ij}a_j$, where $\tilde{A}_0 =  a_3$,
$\tilde{A}_1= -a_2$, $\tilde{A}_2 = a_1$ and $\tilde{A}_3=-a_0$.
$M_{\mu}$ is the chemical potential dependent mass given in
(\ref{mmu}). In this way we get the Hamiltonian

\be H=M_{\mu}-2\lambda \mu^2+\frac{\pi^2_i}{8\lambda},  \ee

\noindent where the canonical momentum is given through a minimal
coupling $\pi_i=p_i-4\lambda\mu \tilde{A}_i$. The Hamiltonian can be
expressed as

\be H = M_{\mu}+\frac{p_i^2}{8\lambda}-\mu \tilde{A}_ip_i, \ee

\noindent and considering the canonical quantization procedure $p_i
\rightarrow \hat{p}_i=-i\delta/\delta a_i$, we get

\bea H=M_{\mu}-\frac{1}{8\lambda}\frac{\delta^2}{\delta a_i^2}-2\mu
\hat{I}_3, \label{hamiltonian}\eea

\noindent where $\hat{I}_3$ is the third component of the isospin
operator \cite{ad-witt}

\be \hat{I}_k= \frac{i}{2}\left(  a_0 \frac{\delta}{\delta a_k} -
a_k \frac{\delta}{\delta a_0} -
\varepsilon_{klm}a_l\frac{\delta}{\delta a_m} \right). \ee

Following the usual procedure, we may associate a wave function to
the Skyrme Hamiltonian. In order to identify baryons in this model,
these wave functions have to be odd, i.e. $\psi(A)=-\psi(-A)$. In
particular, nucleons correspond to linear terms in the $a$'s,
whereas the quartet of $\Delta$'s are given by cubic terms.

\begin{figure}
\includegraphics[angle=0,width=0.5\textwidth]{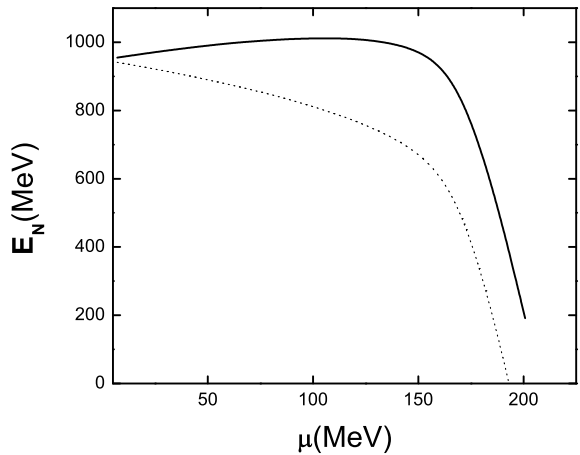}
\caption{\label{mu1mu2} The Nucleon static energy as function of the
isospin chemical potential $\mu$. The splitting between
neutron(solid line) and proton(dotted line) is due to the last term
in (\ref{hamiltonian}).}
\end{figure}

The energy spectra of nucleons as function of $\mu$ is shown in
figure \ref{mu1mu2}. We can see that an energy splitting between
neutrons and protons is induced.

The Skyrmion model allows the existence of different conserved
currents and their respective charges \cite{alvarez}. Using
different charge densities we may define several effective radii.
The evolution of those radii as function of chemical potential
provides information about the critical behavior close to the phase
transition, where the Skyrmion is no longer stable. Let us first
start with the topological baryonic current

\be B^{\mu}=\frac{\varepsilon^{\mu\nu\alpha\beta}}{24 \pi^2} Tr
\left[(U^{\dag}\partial_{\nu}U)(U^{\dag}\partial_{\alpha}U)(U^{\dag}\partial_{\beta}U)
\right]. \label{barionic}\ee

\noindent The baryonic charge density for the Skyrmion is given by

\be \rho_B=4 \pi r^2 B^0(r)=-\frac{2}{\pi}\sin^2F(r) F'(r). \ee

\noindent Obviously, $\int_0^{\infty}dr \rho_B=1$, independently of
the shape of the skyrmionic profile. The isoscalar mean square
radius is defined by

\be \langle r^2 \rangle_{I=0}=\int_0^{\infty}dr r^2 \rho_B.\ee

\begin{figure}
\includegraphics[angle=0,width=0.5\textwidth]{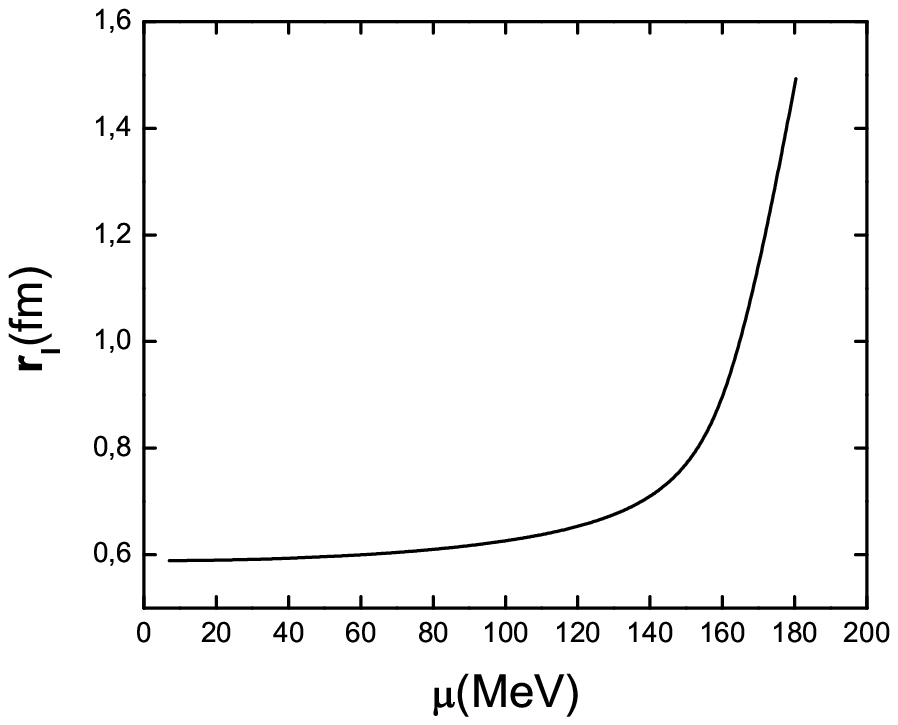}
\caption{\label{risoscalar} The isoscalar mean square radius as
function of $\mu$.}
\end{figure}

This radius seems to be quite stable up to the value of $\mu \approx
120$ MeV, starting then to grow dramatically. Although we do not
have a formal proof  that this radius diverges at a certain critical
$\mu=\mu_c$, the numerical evidence supports such claim, as it is
shown in figure \ref{risoscalar}.

Divergent behavior for several radii, associated to different
currents, has also been observed in different hadronic effective
couplings as function of temperature in the frame of thermal QCD sum
rules \cite{loewe-dominguez}. Similar behavior is found for the mean
square radius associated to the isoscalar magnetic density

\be \rho_M^{I=0}(r)=\frac{r^2F'\sin^2F}{\int dr r^2F'\sin^2F },\ee

The divergent behavior of the the effective radii, suggests the
occurrence of a phase transition, in reference \cite{lmr} the
behavior of the nucleons magnetic moments is also presented, which
also have a divergent behavior.


\bigskip

\noindent
 {\bf Acknowledgements:}   We acknowledge support from Fondecyt
 under grants 1051067 and 1060653

\end{document}